\documentclass[prl,floatfix,twocolumn]{revtex4}
\usepackage{hyperref}

\usepackage{graphicx}
\usepackage{amssymb}

\def\barray{\begin{array}}
\def\earray{\end{array}}
\def\be{\begin{equation}}
\def\ee{\end{equation}}
\def\ben{\begin{equation} \nonumber}
\def\een{\end{equation}}
\def\ban{\begin{eqnarray*}}
\def\ean{\end{eqnarray*}}
\def\ba{\begin{eqnarray}}
\def\ea{\end{eqnarray}}

\def\({\left(}
\def\){\right)}
\def\half{{1\over2}}

\begin{document}

\title{Galilean-invariant scalar fields can strengthen gravitational lensing}
\author{Mark Wyman}
\affiliation{Department of Astronomy and Astrophysics\\
The University of Chicago\\
5640 S Ellis St. \\
Chicago IL 60637, USA}
\email{markwy@oddjob.uchicago.edu}

\begin{abstract}
The mystery of dark energy suggests that there is new gravitational physics on long length scales. Yet light degrees of freedom in gravity are strictly limited by Solar System observations. We can resolve this apparent contradiction by adding a Galilean-invariant scalar field to gravity. Called Galileons, these scalars have strong self-interactions near overdensities, like the Solar System, that suppress their dynamical effect. These nonlinearities are weak on cosmological scales, permitting new physics to operate. In this Letter, we point out that a massive gravity inspired coupling of Galileons to stress energy gravity can have a surprising consequence: enhanced gravitational lensing. Because the enhancement appears at a fixed scaled location for a wide range of dark matter halo masses, stacked cluster analysis of weak lensing data should be able to detect or constrain this effect.
\end{abstract} 
\maketitle


Our understanding of cosmology has been profoundly affected by the discovery of cosmological acceleration. 
It may signal a breakdown of General Relativity on long length scales. This has 
initiated a search for consistent modifications of GR. The leading models for modifying gravity are scalar-tensor theories:
 chameleonic / f(R) theories \cite{cham,fR} and
 the Dvali-Gabadadze-Porrati (DGP) model \cite{DGP}
and its descendants \cite{cascading}. 
Until now, models have assumed that the
scalar couples only to the trace of the stress-energy tensor. Since radiation's stress-energy is trace-free,
gravitational lensing is unaltered. 

The scalar field found in the decoupling limit of the DGP model, $\pi$ \cite{Luty:2003vm}, has an intriguing quality: it is galilean invariant in the action.
That is, the scalar part of the action is unchanged under the replacement $\pi \rightarrow \pi + c + b_\mu x^\mu$,
where $c$ and the $b_\mu$ are arbitrary constants. This galilean symmetry can arise as a manifestation of higher-dimensional 
symmetries \cite{DBIGal}, emerge as a consequence of giving the graviton a mass \cite{massgrav,decoupling, hel0}, 
or simply be posited as a foundation for model building \cite{galileon}.  Fields with galilean-invariant actions are special: they are a
symmetry-protected set of derivatively self-coupled fields with higher-order derivative actions, but with
equations of motion that have only two derivatives operating on the
field at a time.  Equations of motions with more than two time derivatives are in danger of being ill-defined. 
Fields with this symmetry are broadly called {\it galileons}.
Galileon equations of motion contain derivative terms raised to higher powers. 
The non-linearities introduced by these terms
allow galileons to exhibit the Vainshtein mechanism \cite{vainshtein}: the scalar field becomes strongly coupled to itself
near matter sources. This suppresses its gradients. Since the scalar force comes from gradients,
their dynamical influence is suppressed near matter sources. This allows galileons to pass solar system tests
while still having non-trivial effects on longer length scales.  These effects include observables, like extra large
scale structure and faster peculiar velocities \cite{Wyman:2010jp}.

In the decoupling limit of massive gravity \cite{decoupling}, de Rham et al. find a galileon-type theory with an additional
coupling to stress-energy. This has a profound consequence: they are able to degravitate \cite{hel0,degrav},
or suppress the background curvature caused by,  the cosmological constant at the linearized level. In this
letter,
we point out that couplings of the form described in \cite{decoupling} also have a striking phenomenological 
consequence: they can significantly strengthen gravitational lensing
relative to GR.

\begin{figure}[t!] 
   \centering
   \includegraphics[width=0.45\textwidth]{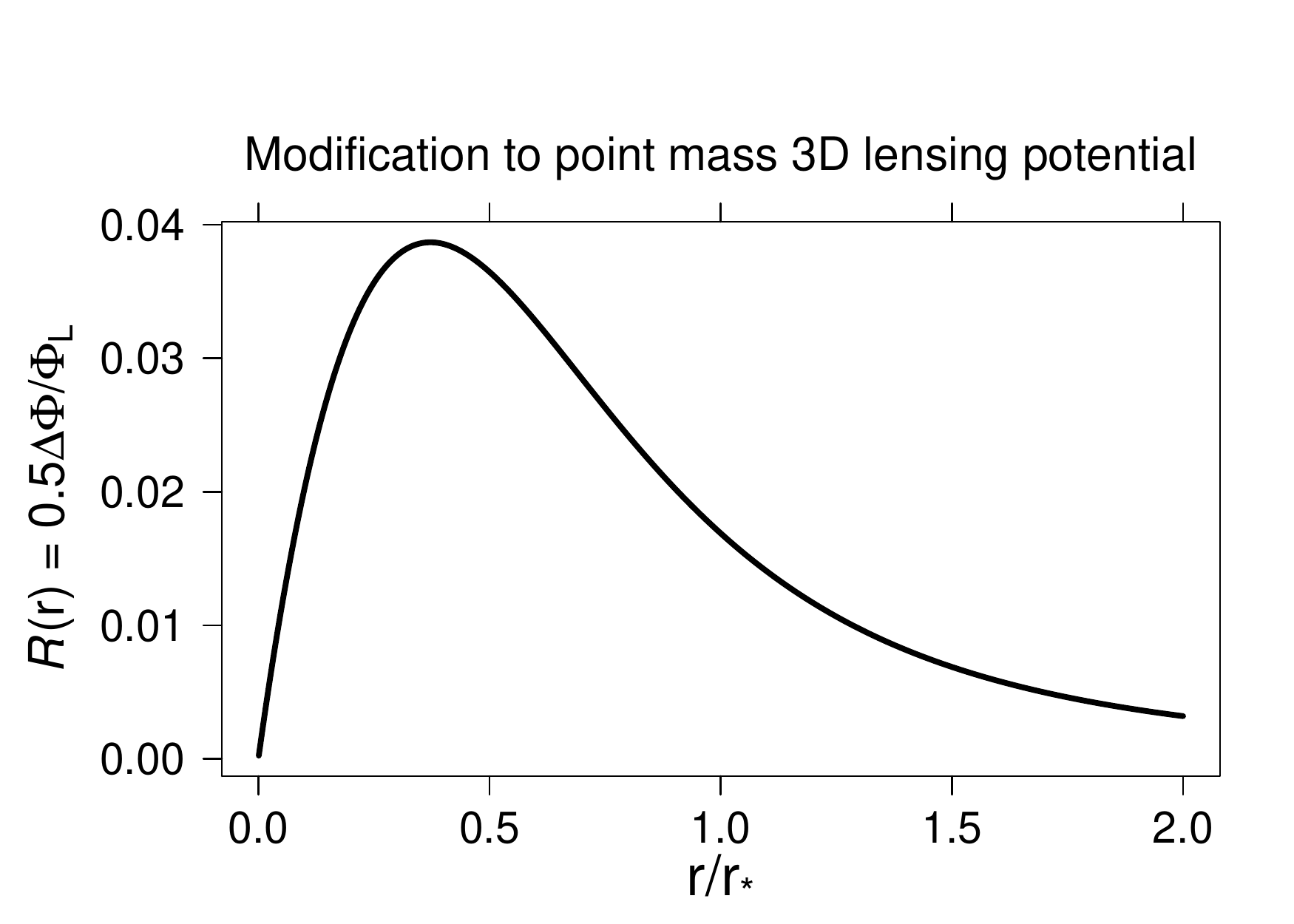} 
   \caption{Radial dependence of the fractional change in the lensing potential, Eqn. \ref{rat}, for a point-like central mass, 
   assuming $a_1=-1/2$ and $a_2=1/2$  (which gives $\alpha=1$, $\beta=1$, $\eta=1$, $\mu=3/2$, and $\nu=1/2$) in the scalar field equations.
   The radius is scaled by $r_* = (r_s r_c^2)^{1/3}$, where $r_s$ is the Schwarzschild radius of the source and $r_c$ is the Compton 
   wavelength (or inverse mass) of the graviton, typically $\sim c/H_0$. For the sun, $r_* \sim$ kpc; for a typical galaxy $r_* \sim$ Mpc; and for a galaxy cluster, 
   $r_* \sim$ 10 Mpc. The peak change of $\sim$4\% is achieved for $r\simeq0.33\, r_*$}
   \label{ratio}
\end{figure}

\begin{figure}[t!] 
   \centering
   \includegraphics[width=0.45\textwidth]{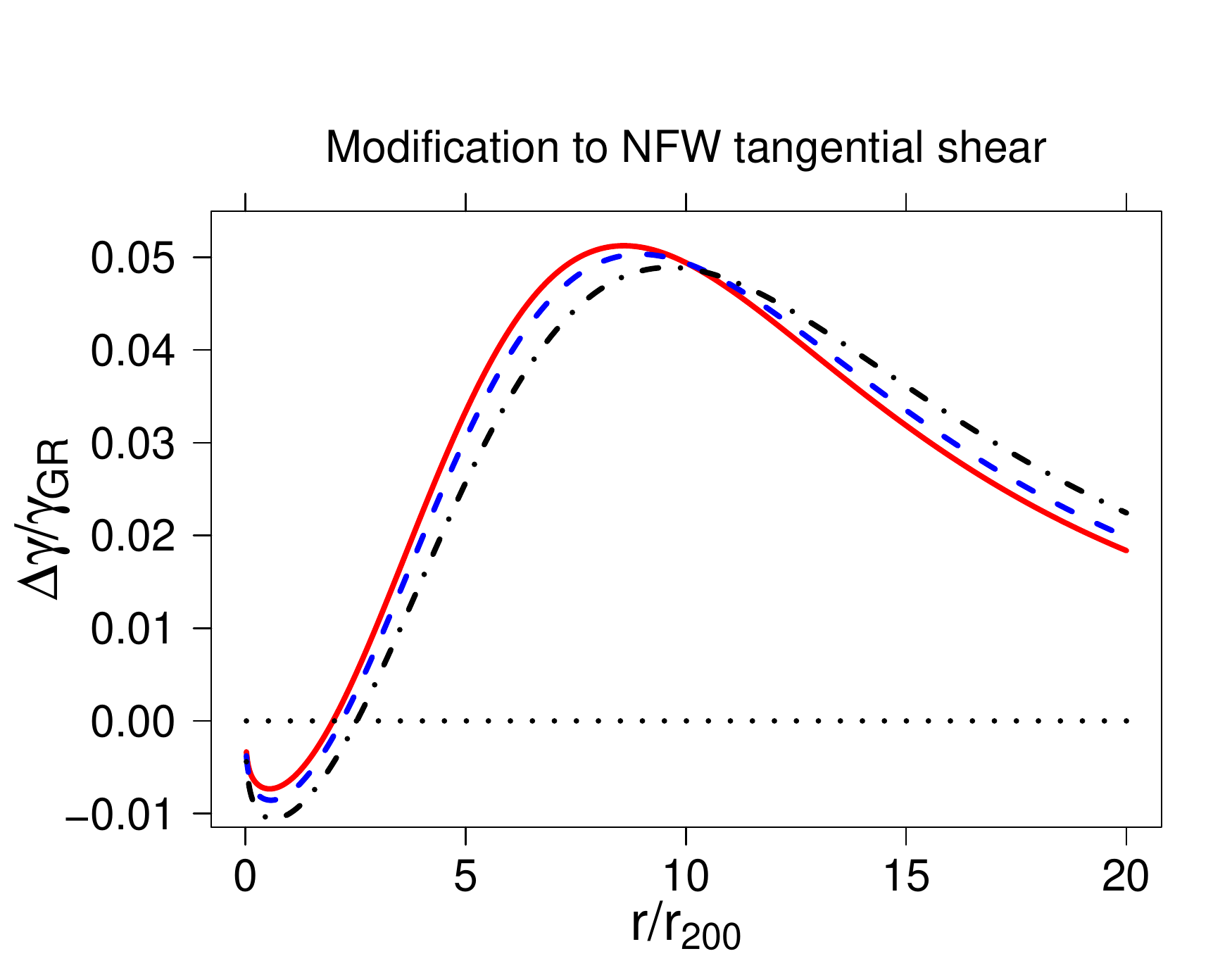} 
   \caption{The fractional change in the tangential shear for three NFW halo profiles as a function
   of radius scaled by the halo's virial radius ($r/r_{200}$),  assuming $a_1=-1/2$, $a_2=1/2$ and $r_c = 3000$ Mpc
   in the scalar field  equations.  These plots describe halos with any virial mass, $M_{200} \propto r_{200}^3$. The three concentrations plotted
   are $c=8$ (solid red), $c=6$ (dashed blue), and $c=4$ (dash-dotted black).  }
   \label{nfw}
\end{figure}

The basic features of this enhancement are as follows. For a spherically symmetric source, it vanishes as $r\rightarrow0$, giving 
negligible Parameterized Post Newtonian (PPN) effects. It also tends to zero as $r\rightarrow \infty$, the limit where the dynamical
effect of the field is largest. For the parameters of the massive gravity model, the lensing
shear is enhanced $\sim5\%$ relative to GR for any spherically symmetric mass configuration.
The increased shear occurs at an intermediate length scale within the strong coupling radius of the theory,
the so-called Vainshtein radius -- see Fig. \ref{ratio}.
This radius is given by $r_*= (r_s r_c^2)^{1/3}$, where
$r_s$ is the Schwarzschild radius of the source and $r_c$ is the Compton wavelength associated with the graviton, typically $\sim c/H_0$.
For the sun, $r_* \sim$ kpc; for a typical galaxy $r_* \sim$ Mpc; and for a galaxy cluster, 
$r_* \sim$ 10 Mpc. In the NFW profile, the 
change in shear is at the percent level for a wide range of radii (Fig. \ref{nfw}). 
This lensing effect is qualitatively different from the parametrized deviations from GR discussed in e.g. \cite{Hu:2007pj}: it is a localized,
inherently nonlinear effect that disappears on long length scales and in linearized perturbation theory. It appears at length
scales that are very well measured by galaxy surveys. The effect is nearly constant in $r_{200}$ units for different halo masses and concentrations. Hence, it should be possible to discover or constrain this effect by stacked analysis of many halos' weak lensing data rescaled by their virial radii.
Planned experiments like the Large Synoptic Survey Telescope (LSST) \cite{LSST} should have sufficient depth to observe this effect. 

{\bf Enhancing the lensing potential:}
For the decoupling limit galileon-type scalar field $\pi$, called the helicity-0 graviton, that arises in theories of massive gravity \cite{decoupling,hel0}, the coupling of the field to stress-energy has the form
\be
\mathcal{L} \subset (h_{\mu \nu} + \alpha \pi \eta_{\mu \nu} + \frac{\beta}{\Lambda_3^3}\partial_\mu \pi \partial_\nu \pi)T^{\mu \nu}, \label{coupling}
\ee
where $\alpha$ and $\beta$ are $\mathcal{O}(1)$ dimensionless coefficients, 
$\Lambda_3 = (M_{Pl} \,m_g^2)^{1/3}$ is the strong coupling scale of the theory, 
$M_{Pl} = (1/G)^{1/2}$ is the Planck mass, and $m_g$ is  the mass of the graviton. For our estimates we will take the graviton
to have a Compton wavelength, $r_c = m_g^{-1} \simeq c/H_0$, the Hubble scale today; and we will work in units where $G=c=\hbar=1$.
 This ``Einstein frame" result is the simplest version of a class of theories studied in \cite{decoupling,hel0}. In this limit, the metric can be
diagonalized and the scalar's effect more easily isolated. Despite these complications, it is clear from the derivations in \cite{decoupling, hel0}
that the metric whose geodesics determine the paths of
photons is the one that includes both tensor ($h_{\mu \nu}$) and scalar ($\partial_\mu \pi \partial_\nu \pi$) parts.
Earlier studies \cite{DGP,galileon} of galileon fields did not contain the coupling $\propto (\partial \pi)^2$ although the absence 
of this coupling means that the $\pi$ field's stress-energy coupling  is not obviously invariant under the galilean symmetry.

As pointed out in \cite{hel0}, this novel coupling permits the degravitation of a small cosmological constant in the decoupling limit. 
In this note, we point out that this coupling has another consequence: the enhancement of the gravitational lensing potential.

For linearized GR, we have $h_{00} =\Psi$, $h_{ij}= \Phi \delta_{ij}$. For lensing in standard GR, the relevant potential is then given by
$\Phi_L = \half(\Phi - \Psi).$
In the presence of a spherically symmetric mass distribution, galileons generically have a non-trivial $\partial_r \pi$ and an approximately
 vanishing $\dot{\pi}$. The additional coupling changes the equations of motion slightly, but $\dot{\pi}\rightarrow0$ is still a good solution. The extra coupling in the lagrangian implies that the potential $\Phi$ is modified, leading to a fractional change $\mathcal{R}(r)$ in the lensing potential $\Phi_L$ given by :
 \be
 \Phi \rightarrow \Phi + \Delta \Phi, \quad  \mathcal{R}(r) \equiv \frac{\half \Delta \Phi}{\Phi_L \mbox{[GR]}}\, ; \quad  \Delta \Phi = \frac{\beta}{\Lambda_3^3}(\partial_r \pi)^2. \label{defR}
 \ee

For our estimates, we will work with a general galileon theory \cite{galileon}, using coefficients consistent with
\cite{decoupling,hel0} and including the extra stress-energy
coupling found in \cite{decoupling,hel0} and given in Eqn. \ref{coupling}. The scalar part of this theory then has the lagrangian

\ba
\mathcal{L}_\pi & = & \frac{3\eta}{2} (\partial \pi)^2 + \frac{\mu}{\Lambda_3^3} (\partial \pi)^2\Box \pi\nonumber + \frac{\nu}{\Lambda_3^6}\([\Pi]^2(\partial \pi)^2 - \right . \\
&& \nonumber \left . 2 [\Pi] \partial_\mu \Pi^{\mu}_\nu \partial^\nu \pi  - [\Pi^2] (\partial \pi)^2 + 2 \partial_\mu \Pi^\mu_\nu \Pi^\nu_\lambda \partial^\lambda \pi\)\\
&& +  (\alpha \pi \eta_{\mu \nu} + \frac{\beta}{\Lambda_3^3}\partial_\mu \pi \partial_\nu \pi)T^{\mu \nu} .
 \ea
In this equation, we have abbreviated some expressions: $(\partial \pi)^2 \equiv \partial_\mu \pi \partial^\mu \pi$ and $\Pi^\mu_\nu \equiv \partial^\mu \partial_\nu \pi$. We have also included five dimensionless $\mathcal{O}(1)$ coefficients, $\alpha$, $\beta$, $\eta$, $\mu$ and $\nu$. 
Although gradients of $\pi$ are suppressed near matter sources by the Vainshtein mechanism, the appearance of the small scale $\Lambda_3^{-3}$ in the gradients' coupling to stress-energy permits $\Delta \Phi$ to become large.
As we will see, the fractional change in $\Phi$ is largest when $\pi' \equiv \partial \pi / \partial r  \propto r^{-1/2}$.
In spherical symmetry, the equation of motion for $\pi$ becomes an
algebraic equation for $\pi'$. This  equation is \cite{galileon} 
\be
3 \eta \(\frac{\pi'}{r}\) + \frac{4 \mu M_P}{\Lambda_3^3} \(\frac{\pi'}{r}\)^2 + \frac{8 \nu M_p^2}{\Lambda_3^6}\(\frac{\pi'}{r}\)^3 = \frac{\alpha GM(r)}{r^3}.
\label{galeom}
\ee
This admits a general closed form solution which is too lengthy to reproduce here.
We have included 5 free coefficients thus far, but in the massive gravity \cite{decoupling} case
these are derived from just two parameters, $a_1$ and $a_2$:
$\alpha =-2 a_1$, $\beta = 2 a_2$, $\eta = 4 a_1^2$, $\mu=-6a_1 a_2$, and $\nu=2 a_2^2$. (In
\cite{hel0}, there is also a third free parameter, $a_3$. When $a_3\neq 0$, the action cannot be diagonalized
into scalar and tensor components. Since this makes the physics
more difficult to understand and is unnecessary to our purposes, we leave $a_3=0$).
This reduction of the parameter space gives a form of the solution for $\pi'(r)$ that
is different and simpler than the general cubic solution, due to a cancellation that occurs when
$2 \mu^2 = 9 \eta \nu$. Note also that  \cite{galileon} finds general constraints on the parameters; for instance, $a_1<0$ is required
for radial perturbative stability.  We will specialize to the $a_1, a_2$ 
parameters for the remainder of this paper. The solution to Eqn. \ref{galeom} as a fraction of the Newtonian force, 
$\Psi'$, is given in terms of $x = r/r_*$, $r_*\equiv(2GM\,r_c^2)^{1/3}$, by
\be
\frac{\pi'}{\Psi'}=x^2 \left [ \( \frac{-a_1}{2\,a_2^2} \)^{1/3} \(  \frac{2a_1^2}{a_2}\, x^3+1 \)^{1/3} + \frac{a_1}{a_2}  x\right ]. \label{piprime} 
\ee

Next, we insert Eqn. \ref{piprime} into  Eqn. \ref{defR} and study its behavior for a point mass.
The first thing to check is that the lensing modification vanishes near the origin, since gravitational lensing in this 
regime is tightly constrained by various PPN tests.  We need
\be
\frac{\partial \pi}{\partial r}(r\sim 0)  \propto r^{n}, \, \quad \quad n>-\half \label{rule}
\ee
so that the behavior of the ratio $\Delta \Phi / \Phi_L \rightarrow 0$ as $r\rightarrow 0$. This is what we find. Interestingly,
the $\nu=0$ case -- which recovers the galileon theory that emerges in the DGP model -- has $n=-1/2$. This implies 
that the modification to lensing from a DGP-like scalar would be non-zero at the origin. Since the enhancement amplitude
is independent of $\Lambda_3$, it persists even in the $m_g \rightarrow \infty$ limit. This is forbidden
by numerous PPN tests of GR. So inclusion
of higher-order terms in the galileon lagrangian was critical for finding an effect that is not already ruled out.
This degree of non-linearity arises naturally in \cite{decoupling}.

For our solution, the behavior near zero is given by
\be
\frac{\partial \pi}{\partial r}(r\sim 0)  = \frac{r_*}{2\sqrt[3]{2}\, r_c^2} \(\frac{a_1}{a_2^2}\)^{1/3}  + \frac{r}{2 \, r_c^2} \frac{a_1}{a_2} + \mathcal{O}(r^3),
\ee
i.e., approaching a small constant near $r=0$, giving an $n=0$ scaling in Eqn. \ref{rule}. Thus our solution does not violate
solar system tests.

The other limit to check is $r\rightarrow \infty$. Here again, the ratio vanishes, since
galileon theories generically recover $\pi'(r\rightarrow\infty) \propto 1/r^2$, so it scales as $1/r^3$ for large $r$.

These limiting behaviors imply that the solution must at some point pass through the $r^{-1/2}$ scaling that will give
a $\Delta \Phi$ with the same radial scaling as $\Phi_L[\mbox{GR}]$ and hence a finite rescaling of the strength
of gravitational lensing. For the parameters of the massive gravity model  and general $r$, the fractional change in the 3D lensing potential
is given by
\be
\mathcal{R}(x) = \frac{x}{8\,a_2} \((-4\,a_1a_2)^{1/3}\(\frac{2\,a_1^2}{a_2} x^3+1\)^{1/3} +  2\,a_1x\)^{2} \label{rat},
\ee
where $x=r/r_*$, $r_* \equiv (2 G M r_c^2)^{1/3}$. Note that the ratio takes 
a particularly simple form for $a_1=-1/2$, $a_2=1/2$; we will make this choice in our plots. 
This choice also gives the same long-distance dynamics as the DGP model (i.e., $\pi'(r) / \Psi_N'(r) \rightarrow 1/3$
for $r\gg r_*$).
We have plotted $\mathcal{R}(r)$ in Fig. \ref{ratio}.
$\mathcal{R}(r)$ reaches
a maximum at $x_o=r_o/r_*= ((2\sqrt{3}-3)a_2/18a_1^2)^{1/3}$ given by
\be
\mathcal{R}(x_o) = \frac{1}{12}(2\sqrt{3}-3) \simeq 0.04.
\ee
This peak amplitude is independent of the parameter choices $a_1$ and $a_2$ and can be regarded as a prediction of the theory.
Though small,  this modification gives a potentially observable modification to the tangential shear of extended halos; this is illustrated in Fig. \ref{nfw}.

{\bf Weak lensing:} 
The enhancement to lensing we are studying peaks on intermediate length scales. 
Weak lensing around galaxies and clusters is thus the best place to look for its effects.
Hence, we calculate the effective change in the
tangential lensing shear caused by the galileon for a Navarro-Frenk-White (NFW) halo profile, following \cite{wrightbrainerd}. We plot this 
for three different halo concentrations in Fig. \ref{nfw}. (N.B. Existing parameterizations of
modified gravity (e.g. \cite{Hu:2007pj}) are designed to work on scales
characterized by linear overdensities, $k \lesssim 0.1$ h/Mpc.  The effects we are describing
vanish on those scales, so they are not adequate to studying this effect.)

For an NFW halo, the modification peaks at $\sim0.5\, r_{200}$ and at $\sim 9 \, r_{200}$ for this parameter set, where $r_{200}$ 
is the virial radius. It depends quite weakly on halo concentration. 
Because $M_{200} \propto r_{200}^3$ and $r_* \propto M_{200}^{1/3}$, the 
effect peaks at the same locations, as measured in units of $r_{200}$, for all $M_{200}$. This
 makes the effect potentially observable: we can stack the lensing results from many clusters, scaled
by their virial radii, and look for the effect to emerge statistically. The same reasoning also implies that
 the character of the modification will be redshift independent if the galileon's parameters
 do not depend strongly on cosmology. We should caution that this cosmological behavior is not well understood.
  A simple estimate of when the effect turns on is when the Universe comes within its 
own $r_*$, which occurs around $z\sim1$. So our predictions are likely robust for $1 \gtrsim z  \ge 0$.

{\bf Detectability:} Over the easily observable range $r<1.5 r_{200}$, $\langle | \Delta \gamma |\rangle \sim 1\%$ (Fig. \ref{nfw}). 
Taking this as a signal above a known background, we can estimate
what observations are needed to detect it.
The GR shear at these radii is $\gamma\sim 10^{-2}$.
Assuming a shape variability of $\sigma_\gamma =0.3$, we find 
$N_{obs}\sim10^7$ observations are needed for $S/N \gtrsim 1$.  We can estimate
$N_{obs}\simeq(N_{gal}/\mbox{arcmin}^2) N_{\mbox{lenses}} A_{\mbox{lens}}$.
We can get $N_{obs}\sim10^7$ with an LSST-like depth of $40$ galaxies / arcmin$^2$ \cite{LSST} if we stack, e.g., $>5\times10^4$ 
lenses that each subtend 5 arcmin$^2$. 
Our data cannot achieve this \cite{Reyes:2010tr}. We are
performing a more thorough study of detectability now \cite{tocome}.

{\bf Strong lensing:} To see the effect of the galileon coupling on strong lensing, we can find a solution to Eqn. \ref{galeom}
for a singular isothermal sphere (SIS) and study its behavior near $r=0$, the strong lensing regime.
 The SIS has a density $\rho(r) \propto r^{-2}$ and a mass profile
$\propto r$. It turns out that the galileon-sourced 2D shear profile can be calculated in closed form for the SIS in terms of hypergeometric functions.
For the SIS, the galileon-generated fractional increase in the lensing potential
grows as $(r/r_c)^{2/3}$ for small $r$. This means that the galileon field generates an effective projected
density profile $\Sigma (\xi\sim0) \propto \xi^{-1/3}$, where $\xi$ is the 2D radial distance from the center
of the source after the line-of-sight direction has been integrated out. Unfortunately, this component is quite small.
For this additional source of effective surface density to generate even a $>1$\% increase in the effective
Einstein radius, $\theta_E$, the mass per radius of the SIS would have to be $\gtrsim 10^{13} M_\odot/$Mpc. 
This is unlikely to account for the apparent excess of lensing arcs
seen in gravitational lensing surveys, e.g. \cite{lensing}.

{\bf Conclusions:}
In this paper, we have given a first study of the modifications to gravitational lensing generated by the 
inclusion of a new coupling of a scalar component of gravity to stress-energy. This coupling
arises naturally in ghost-free theories of massive gravity \cite{decoupling}, and is reasonable to include
in phenomenological theories of galilean-invariant scalar fields. The generic effect of this coupling is to
strengthen gravitational lensing on length scales $\sim 0.5\, r_*$, where  
$r_* = (r_s r_c^2)^{1/3}$; $r_s$ is the Schwarzschild radius of the source and $r_c$ is the Compton 
wavelength (or inverse mass) of the graviton, typically $\sim c/H_0$. For the sun, $r_* \sim$ kpc; for a typical galaxy $r_* \sim$ Mpc; and for a galaxy cluster, 
$r_* \sim$ 10 Mpc.
The enhancement to tangential shear is at the percent level for the parameter combinations that appear in the massive
graviton version of the galileon theory. The enhancement appears at a fixed location in relation to a halo's virial radius
for a wide range of masses and concentrations. 
This should allow stacked analysis of weak lensing data to 
measure or constrain this effect. 
   
{\bf Acknowledgements:} We are very grateful to Neal Dalal, Claudia de Rham,
Mike Gladders, Wayne Hu, and Melanie Simet for extended discussions, to the anonymous referees for
constructive suggestions, and
to Sasha Belikov, Will High, Rachel Mandelbaum, Beth Reid, Youngsoo Park, Ali Vanderveld, and Dan Wesley 
for helpful exchanges. The work of M.W. at the University of Chicago is supported by the Department of Energy.

\end{document}